\documentstyle[12pt,jour]{article}
\def\sbox#1{\mbox{\small #1}}
\def\Tr{\mbox{Tr}}
\def\H{{\cal H}}
\begin{document}
\large
\sloppy
\title{\bf Magnetic Properties of the Spin-1/2
Ferromagnetic-Ferromagnetic-Antiferromagnetic Trimerized Heisenberg Chain}
\vspace{5mm}
\author{\bf Kazuo Hida \\ {\it Department of Physics,}
\\ {\it College of Liberal Arts,} \\ {\it Saitama University,} \\
 {\it Urawa, Saitama 338} \\ {(Received\hspace{7cm})}}
\date{}
\maketitle
\baselineskip=8.5mm
{\noindent{\bf Abstract}}

\vspace{5mm}
The magnetic properties of the ferromagnetic-ferromagnetic-antiferromagnetic
trimerized spin-1/2 Heisenberg chain are studied theoretically. The high
temperature susceptibilty and the ground state saturation magnetic field are
calculated and the exchange energies of the trimer compound 3CuCl${}_2\cdot$2dx
are determined. The magnetization curve is obtained by numerical
diagonalization of finite size systems. The result explains the low temperature
magnetization data for 3CuCl${}_2\cdot$2dx with the exchange energies obtained
as above. It is predicted that the magnetization curve has a plateau at 1/3 of
the saturation magnetization if the ferromagnetic exchange energy is comparable
to or smaller than the antiferromagnetic exchange energy.

\vspace{2mm}

\noindent
e-mail: hida@th.phy.saitama-u.ac.jp

\newpage
\section{Introduction}
Recently, the quantum composite spin systems with mixed interactions have been
studied extensively.\cite[-14]{kh1} For example, the
ferromagnetic-antiferromagnetic alternating chain\cite[-3]{kh1} and the
Heisenberg ladder model\cite[-8]{kh2} provided the key idea to understand the
Haldane phase of the spin-1 uniform Heisenberg chain. The latter is also
relevant for the understanding of the supercondutivity in CuO
compounds\cite[-8]{drs1}. Two dimensional antiferromagnetic Heisenberg models
with alternating bonds\cite{ki1} and double layer geometry\cite{kh3} have been
also studied in relation to the mechanism of superconductivity. The interacting
trimer models have been studied by Zaspel\cite{cz1,cz2} using the expansion
from the independent trimer limit. The trimerized spin-1/2 XY model has been
also studied by Okamoto.\cite{ko1}

In the present work, we study the spin-1/2
ferromagnetic-ferromagnetic-antiferromagnetic Heisenberg chain which can be
regarded as one of the interacting trimer models. This model is expected to
behave similarly to the spin-3/2 Heisenberg antiferromagnetic chain for strong
ferromagnetic coupling. However, for small ferromagnetic coupling, the quantum
fluctuation is enhanced and wider variaty of phenomena should be observed. Even
for the moderate ferromagnetic coupling, substantial correction to the $S=3/2$
chain is expected.

It should be emphasized that this model is not merely a theoretical toy.
Livermore et al.\cite{liv1} measured the magnetic susceptibility of the
compound 3CuCl${}_2\cdot$2dx (dx = Dioxane) which consists of
antiferromagnetically coupled ferromagnetic trimers. Recently, Ajiro et
al.\cite{ajiro1} have measured  the magnetic susceptibility and the
magnetization curve of this material at low tempratures.

This paper is organized as follows. In the next section, the model Hamiltonian
is defined. The result of the high temperature expansion for the susceptibility
is presented in section 3 and the ground state saturation field is analytically
obtained in section 4. The exchange energies of 3CuCl${}_2\cdot$2dx are
determined by fitting the high temperature susceptibility and the saturation
field to the formula obtained in sections 3 and 4. In sectin 5, the ground
state magnetization curve is determined by numerical diagonalization of finite
size systems and compared with the experiment in for 3CuCl${}_2\cdot$dx The
last section is devoted to summary and discussion.

\section{Ferromagnetic-Ferromagnetic-Antiferromagnetic Trimerized Heisenberg
Model}

Our model is described by the Hamiltonian:
\begin{equation}
\H=\sum_{i=1}^{N} \{\H_i^{\sbox{trim}} + \H_i^{\sbox{int}}\},
\end{equation}
where
\begin{eqnarray}
\label{eq:haml}
\H_i^{\sbox{trim}} &=&-2J_F( \v{L}_i \v{M}_{i} + \v{M}_i \v{R}_i), \\
\H_i^{\sbox{int}} &=& 2J_A \v{R}_i \v{L}_{i+1},
\end{eqnarray}
Here $J_A$ and $J_F$ are the antiferromagnetic and ferromagnetic couplings,
respectively. The operators $\v{L_i}$, $\v{M_i}$ and $\v{R_i}$ are $S=1/2$ spin
operators. In the following, the ratio of the two exchange couplings is denoted
by $\gamma = J_F/J_A$. In the limit $\gamma \rightarrow \infty$, the spins
$\v{L}_i$, $\v{M}_i$ and $\v{R}_i$ form a single spin $\v{S}_i$  with magnitude
3/2 and this model reduces to the spin-3/2 antiferromagnetic Heisenberg chain.
The exchange energy of the resulting spin-3/2 chain can be determined as
follows:

The Hamiltonian $\H_i^{\sbox{trim}}$ can be rewritten as,
\begin{equation}
\H_i^{\sbox{trim}} = J_A\gamma\sum_{\alpha = x,y,z}\left\{
(L^{\alpha}_{i}-M^{\alpha}_{i})^2 + (M^{\alpha}_{i} - R^{\alpha}_{i})^2 - 3
\right\}
\end{equation}
In the limit $\gamma \rightarrow \infty$, therefore, the Hilbert space is
restricted by the following condition.
\begin{equation}
\label{eq:cond}
L^{\alpha}_{i}\mid \Phi >=M^{\alpha}_{i}\mid \Phi >= R^{\alpha}_{i}\mid \Phi >,
(\alpha = x,y,z ; i = 1,..,N)
\end{equation}
Thus, we may replace $\v{R}_i$ and $\v{L}_i$ by $\v{S}_i \equiv (\v{L}_i +
\v{M}_i + \v{R}_i)/3$ in $\H_i^{\sbox{int}}$ as,
\begin{equation}
\H_i^{\sbox{int}} = \frac{2J_A}{9} \v{S}_i\v{S}_{i+1},
\end{equation}
where $\v{S}_i$ can be regarded as a spin-3/2 operator in this restricted
subspace. Thus the effective coupling $J_{3/2}$ for the corresponding spin-3/2
chain is given by $J_A/9$.

\section{Saturation Field}
\vspace{5mm}

The effect of the magnetic field is expressed by the Zeeman term
$\H^{\sbox{Z}}$ in the Hamiltonian as follows,
\begin{equation}
\H^{\sbox{Z}} = - g\mu_B H \sum_{i=1}^{N}(L_i^z + M_i^z + R_i^z)\equiv - m_zH
\end{equation}
where $g, \mu_B$, $m_z$ and $H$ are the $g$-factor of electrons, Bohr magneton,
the total magnetization and the magnetic field, respectively. The ground state
saturation field is given by the field at which the Zeeman energy compensates
the energy difference between the fully magnetized state $\mid 3N/2>_0$ and the
lowest energy state which has a single inverted spin $\mid 3N/2-1>_0$. Here and
in what follows, $\mid S_{\sbox{tot}}^{z}>_0$ denotes the lowest energy state
with $z$-component of the total spin $S_{\sbox{tot}}^{z}$. The energy of the
state $\mid 3N/2>_0$ is simply given by $E_{\sbox{sat}}=N(4J_F - 2J_A)/4$. We
denote the energy of the state $\mid 3N/2-1>_0$ measured from $E_{\sbox{sat}}$
by $\Delta E$.

The state $\mid k >$ with  $S_{\sbox{tot}}^{z}=3N/2-1$ and momentum $k$ is
expanded as,

\begin{equation}
\mid k > = \frac{1}{\sqrt{3N}}\sum_{j=1}^{N}\{ A_1 \mid 3j-2> + A_2 \mid 3j-1>
+ A_3 \mid 3j>\}\exp(3\img jk)
\end{equation}
where $\mid j >$ is the state with a down spin on the $j$-th site and up spins
on other sites. The eigenvalue equation yields the $3 \times 3$-matrix
equation,
\begin{equation}
\left[
\begin{array}{ccc}
J_F -J_A -\Delta E  &  -J_F  & J_A \mbox{e}^{-3\img k} \\
-J_F  &  2J_F - \Delta E   & -J_F \\
J_A \mbox{e}^{3\img k}  &  -J_F  & J_F -J_A -\Delta E
\end{array}
\right]
\left[
\begin{array}{c}
A_1 \\
A_2 \\
A_3
\end{array}
\right]
= 0
\end{equation}
Solving this eigenvalue problem, the saturation field $H_s$ is obtained from
the lowest eigenvalue with $k = \frac{\pi}{3}$ as,
\begin{equation}
\label{eq:de}
g\mu_B H_s = -\Delta E = \sqrt{\left(J_A + \frac{J_F}{2}\right)^2 + 2J_F^2 } -
\frac{3}{2}J_F + J_A
\end{equation}
In the limit $J_F \rightarrow \infty$, $H_s$ tends to $4J_A/3g\mu_B$. This
coincides with the result of Parkinson and Bonner\cite{pb1} with the
identification $J_{3/2} = J_A/9$

\section{High Temperature Expansion}

The magnetic susceptibilty $\chi(T)$ at temperature $T$ can be calculated by
the fluctuation-disspation theorm as,
\begin{eqnarray}
\chi(T) &=& \beta\frac{\Tr m_z^2 \exp(-\beta \H)}{Z} \\
Z &=& \Tr \exp(-\beta \H), \mbox{\hspace{5mm}}\beta =1/T
\end{eqnarray}
where the commutativity of $m_z$ with $\H$ is taken into account. Expanding in
powers of $\beta$, we have,
\begin{equation}
\chi(T)T = \frac{\Tr (m_z^2) - \beta \Tr (m_z^2 \H) + \frac{1}{2} \beta^2 \Tr
(m_z^2 \H^2)+ ...}{\Tr 1 - \beta \Tr (\H) + \frac{1}{2} \beta^2 \Tr (\H^2)+
...}\end{equation}
Collecting the terms up to the second order in  $\beta$, we have,
\begin{equation}
\label{eq:kai}
\chi(T)T \simeq \frac{3N(g\mu_B)^2}{4}\left(1-\frac{J_A - 2J_F}{3T} -
\frac{(J_A +J_F)^2}{6T^2} ....\right)
\end{equation}
Fitting the data of Livermore et al.\cite{liv1} at 200K $\leq T \leq$ 300K to
(\ref{eq:kai}), we find $g \simeq 2.13, J_F \simeq 68$K and $J_A \simeq$ 15.3K.
The values of $J_A$ and $J_F$ are supposed to satisfy (\ref{eq:de}) with the
saturation field $H_s \simeq$ 15T\cite{ajiro1}. Although Livermore et
al.\cite{liv1} estimated as $J_F \simeq$ 94K and $J_A \simeq$ 7.6K, this
estimation is based on the erroneous identification of $J_A$ with  $5J_{3/2}$.
The correct relation should be  $J_A = 9J_{3/2}$ as discussed in section 2.
They also used the crude approximation to replace the Heisenberg type
interaction by the Ising type one.

Zaspel\cite{cz1} performed the expansion in $\beta J_A$ and determined both
exchange energies from the temperature dependence of susceptibility. Although
his expression for the susceptibility contains some errors, he obtained $J_F
\simeq$ 85K and $J_A \simeq$ 14K which are close to our estimation.

\section{Numerical Diagonalization}

We diagonalize the finite-size Hamiltonian (\ref{eq:haml}) with periodic
boundary condition using TITPACK version 2 supplied by Nishimori for
$m_z/g\mu_B = 0, ..., 3N/2$. The ground state magnetization curve is obtained
from the magnetization of the lowest energy eigenstate of $\H + \H^{\sbox{Z}}$
for each value of the magnetic field.

The results are shown in Fig. 1(a)-(d) for $N=4,6$ and 8 with $\gamma = 1.0,
2.0, 5.0$ and 10.0 . The magnetization is normalized by the saturation
magnetization $m_s \equiv 3Ng\mu_B/2$. The solid line is the interpolation
curve of the midpoints of the vertical and horizontal steps. It should be
remarked that the magnetization curve has a plateau at $m_z/m_s = 1/3$ for
small $\gamma$. For these cases(Fig. 1(a) and (b)), the interpolation is made
for the portions $m_z/m_s > 1/3$ and $m_z/m_s < 1/3$ separately. We denote the
lower and upper boundary of the plateau by $H_{c1}$ and $H_{c2}$, respectively.
They are shown in Fig. 2 for finite size systems. The values extrapolated to $N
\rightarrow \infty$ are also shown. The extrapolation is made by fitting the
data for $N=4, 6$ and 8 to the second order polynominal of $1/N$ naively.

For small $J_F$, this behavior is naturally explained as follows: The spins
$\v{L}_i$ and $\v{R}_{i+1}$ connected by the antiferromagnetic bonds (hereafter
called side-spins) tend to form singlet pairs. The remaining $N$ spins
($\v{M}_i$'s; hereafter called mid-spins) are weakly coupled with each other
mediated by the neighbouring side-spins. The effective Hamiltonian for the
mid-spins is obtained by the elementary perturbation theory in
$\H^{\sbox{trim}}$ as
\begin{equation}
\label{eq:eff}
\H^{\sbox{mid}} =\frac{J_F^2}{J_A} \sum_{i=1}^N \v{M}_i
\v{M}_{i+1},\end{equation}
within the second order of $J_F$. Therefore the lower critical field $H_{c1}$
is given by the saturation field of the spin-1/2 uniform Hamiltonian
$\H^{\sbox{mid}}$ as,\cite{pb1}
\begin{equation}
\label{eq:hc1}
g\mu_BH_{c1} \simeq 2J_F^2/J_A = 2J_A\gamma^2
\end{equation}
where all mid-spins are aligned to give $m_z= m_s/3$. This is the state $\mid
N/2>_0$.

This plateau extends up to the upper critical field $H_{c2}$ determined from
the energy difference between the states $\mid N/2>_0$ and $\mid N/2+1>_0$ in
which all mid-spins are aligned, one of the pairs of side-spins form a triplet
and all other side-spins form singlets. For $J_F=0$, this energy difference is
given by $2J_A$. Again using the elementary perturbation theory, the first
order correction to the energy is given by $-J_F$ for $\mid N/2+1>_0$, while it
vanishes for $\mid N/2>_0$. Therefore we have
\begin{equation}
\label{eq:hc2}
g\mu_B H_{c2} \simeq 2J_A-J_F = 2J_A(1-\gamma/2).
\end{equation}
The dashed lines in Fig.2 show the relations (\ref{eq:hc1}) and (\ref{eq:hc2}).

Although the above picture applies only for the weak ferromagnetic coupling,
the numerical data suggest that the plateau clearly exists even for $\gamma
\sim 2$. This is due to the fact that the effective strength of the
ferromagnetic bond is weaker than that of the antiferromagnetic bond with the
same exchange energy, because the ferromagnetic bond gains only $J_F/2$ by
forming a triplet state while the antiferromagnetic bond gains $3J_A/2$ by
forming a singlet state.

The plateau appears to ends up at finite critical value of $\gamma = \gamma_c
\simeq 2 \sim 3$. If this is the case, the transition between the spin-gap
state and the spin liquid state takes place at $\gamma = \gamma_c$ for $m_z
=m_s/3$. From the universality consideration, however, it is also possible to
expect that the plateau persists to $\gamma \rightarrow \infty$ with
exponentially small width, because the symmetry of the system does not change
over the interval $0 < \gamma < \infty$. Within our numerical data, this
possibility cannot be ruled out.

For large enough value of $\gamma$, the magnetization curves are similar to
that of $S=3/2$ chain obtained by Parkinson and Bonner\cite{pb1} with
increasing downward curvature as $\gamma$ decreases. In Fig. 3, the
experimentally observed magnetization curve of 3CuCl${}_2\cdot$2dx at 1.5K is
compared with our numerical calculation with $\gamma=4.5$, because  $J_F
\simeq$ 68K and $J_A \simeq$ 15.3K for this material as discussed in the
preceding sections. The magnetization curves of the uniform spin-1/2 and 3/2
Heisenberg antiferromagnetic chains\cite{pb1} are also shown for comparison.
Obviously, our data are better fitted to the experimental data than those for
the uniform chains.

\section{Summary and Discussion}

The high temperature magnetic susceptibility and the ground state magnetization
curve of the spin-1/2 ferromagnetic-ferromagnetic-antiferromagnetic trimerized
chain is calculated. The exchange energies of the compound 3CuCl${}_2\cdot$2dx
are determined by comparing our calculation of the high temperature magnetic
susceptibility and the ground state saturation field with the experimental
data.

The ground state magnetization curve is obtained by numerical diagonalization
of finite size systems. The result also explains the low temperature
magnetization curve of 3CuCl${}_2\cdot$2dx fairly well using the values of
exchange energies obtained above. It should be noted, however, that the
estimation of $J_F$ is still ambiguous. For example, if we fix the value of $g$
to 2.093 as obtained from the EPR measurement\cite{liv1}, the best fit of the
susceptibility to the high temperature expansion is obtained with $J_F \simeq$
85K and $J_A \simeq$ 15.2K. However, the strong downward curvature of the
experimentally observed magnetization curve favors the small value of $J_F
\simeq 4 \sim 5 J_A$.

The magnetization curve has a plateau at 1/3 of the full magnetization if the
intratrimer ferromagnetic coupling is not much larger than the intertrimer
antiferromagnetic coupling. Therefore it would be interesting to synthesize the
similar compound with smaller values of $\gamma$. From theoretical point of
view, it is also an interesting problem whether this plateau persists to
$\gamma \rightarrow \infty$.

Ajiro et al.\cite{ajiro1} found a weak anomaly in the differential
susceptibility of 3CuCl${}_2\cdot$2dx at $H = 3.5$T. They attributed this
anomaly to the spin-flop type transition. However, it might be also possible to
interprete this anomaly as the precursor of the above mentioned plateau,
although our finite size calculation cannot pick up such a fine structure of
the magnetization curve. This point requires further study.
\vspace{5mm}

\noindent{\bf Acknowledgements}
\vspace{5mm}

The author is grateful to  Y. Ajiro for showing the experimental data of his
group prior to publication and for stimulating discussion. He is also indebted
to H. Nishimori for providing TITPACK version 2. This work is supported by the
Grant-in-Aid for Scientific Research on Priority Areas "Computational Physics
as a New Frontier in Condensed Matter Research" from the Ministry of Education,
Science and Culture. The numerical calculation is performed by HITAC S3800/480
at the Computer Center of the University of Tokyo and HITAC S820/15 at the
Information Processing Center of Saitama University.

\newpage
\noindent{\bf Figure Captions}
\vspace{5mm}
\begin{enumerate}

\item The magnetization curve with $\gamma =$ (a)$1.0$, (b) 2.0, (c) 5.0  and
(d) 10.0 for $N=4$($\circ$), 6($\bullet$) and 8($\Box$). The dotted line is the
interpolation curve of the midpoints of the horizontal and vertical steps of
the data for $N=8$.

\item The lower and upper boundaries of the magnetization step at $m_z=m_s/3$
for $N=4$($\circ$), 6($\triangle$) and 8($\Box$). The filled circles show the
value extrapolated to $N \rightarrow \infty$. The dashed lines are the
approximate relation (\ref{eq:hc1}) and (\ref{eq:hc2}) for small $\gamma$.

\item The magnetization curve of 3CuCl${}_2\cdot$dx at $T$ = 1.5K(solid
line)\cite{ajiro1} and  theoretically calculated magnetization curve with
$\gamma = 4.5$(dotted line). The latter is the interpolation of the midpoints
of the horizontal and vertical steps of the finte system with $N=8$ shown only
for the magnetic field larger than the first step. The magnetization curve of
$S=1/2$(dashed line) and $S=3/2$(dash-dotted line) uniform Heisenberg chain
calculated by Parkinson and Bonner\cite{pb1} are also shown for comparison.

\end{enumerate}

\newpage

\noindent{\bf References}
\vspace{5mm}

\begin{thebibliography}{99}
\bibitem{kh1}   K. Hida: J. Phys. Soc. Jpn. {\bf 62} (1993) 1466 and refernces
therein.
\bibitem{kt1}   M. Kohmoto and H. Tasaki:  Phys. Rev. {\bf B46} (1992) 3486.
\bibitem{yhk1}  M. Yamanaka, Y. Hatsugai and M. Kohmoto: Phys. Rev. {\bf B48}
(1993) 9555.
\bibitem{kh2}   K. Hida: J. Phys. Soc. Jpn. {\bf 60} (1991) 1347; ibid. 1939.
\bibitem{wnt1}  H. Watanabe, K. Nomura and S. Takada: J. Phys. Soc. Jpn. {\bf
62} (1993) 2845.
\bibitem{drs1}  E. Dagotto, J. Riera and D. Scalapino: Phys. Rev. {\bf B45}
(1992) 5744.
\bibitem{bdrs1}  T. Barnes, E. Dagotto, J. Riera and E.S. Swanson: Phys. Rev.
{\bf B47} (1993) 3196.
\bibitem{grs1}  Sudha Gopalan, T.M. Rice and M. Sigrist: preprint (1993).
\bibitem{ki1}   N. Kato and M. Imada: J. Phys. Soc. Jpn. {\bf 62} (1993) 3728.
\bibitem{kh3}   K. Hida: J. Phys. Soc. Jpn. {\bf 61} (1992) 1013 and refernces
therein.
\bibitem{cz1} C.E Zaspel: Phys. Rev. {\bf B39} (1989) 2597.
\bibitem{cz2} C.E Zaspel: J. Magn. Magn. Mater. {\bf 87} (1990) 90.
\bibitem{ko1} K. Okamoto: Solid State Commun. {\bf 83} (1992) 1039.
\bibitem{hs1}	H.J. Schulz: Phys. Rev. {\bf B34} 6372 (1986).
\bibitem{liv1}  J. C. Livermore, R.D. Willett, R.M. Gaura and C.P.Landee:
Inorg. Chem. {\bf 21} (1982) 1403.
\bibitem{ajiro1} Y. Ajiro, T. Asano, T. Inami, H. Aruga-Katori and T. Goto:
preprint(1993) and private communication.
\bibitem{pb1}  J.B. Parkinson and J.C. Bonner: Phys. Rev. {\bf B32} (1985)4703.
\end{thebibliography}
\end{document}